\documentstyle[12pt]{article}
\newcommand{\bq}{\begin{equation}}
\newcommand{\eq}{\end{equation}}
\newcommand{\ds}{\displaystyle}
\newcommand{\lng}{\ln g}
\newcommand{\td}{\tilde{d}}
\begin{document}    \begin{flushright} PITHA 98/10 \\ hep-th/9803180
\end{flushright}
   \begin{center} {\Large    Schwarzschild  Black Hole
  Quantum Statistics,
  \\[0.2cm] Droplet Nucleation and DLCQ Matrix Theory \\[1,5cm] }
{\large \ H.A. Kastrup\footnote{E-Mail: kastrup@physik.rwth-aachen.de} \\
Institute for Theoretical Physics, RWTH Aachen \\[0.2cm] 52056 Aachen,
 Germany}
\end{center} \vspace*{1.0cm}
  {\large  Abstract} \\[0.2cm]
   Generalizing previous quantum gravity results for
    Schwarzschild black holes from 4
    to $D\geq4$ spacetime dimensions yields an energy
   spectrum \[E_n = n^{ 1-1/(D-2)}\, \sigma \,E_{P,D}~,~n=1,2, \ldots~, \sigma =
   O(1)~. \]  Assuming the degeneracies
   $d_n$ of these levels to be given by $d_n=g^n, g>1,$ leads to a partition
   function which is the same as that of the primitive droplet nucleation
   model for 1st-order phase transitions in D-2 spatial dimensions. \\
    Exploiting the well-known
   properties of the so-called critical droplets of this model immediately
   leads to the Hawking temperature and the Bekenstein-Hawking entropy of
   Schwarzschild black holes. Thus, the "ho\-lo\-gra\-phic prin\-ciple"
 of 't Hooft
   and Susskind is naturally
   realised. The values of temperature and entropy  appear closely
    related to the
    imaginary part of the partition function which describes metastable
     states.
     \\
   Finally some striking conceptual similarities ("correspondence
    point" etc.)
    between the droplet
   nucleation picture and the very recent approach to the quantum statistics
   of Schwarzschild black holes in the framework of the DLCQ Matrix theory
   are pointed out.
  \newpage
   \section{Introduction}
    In two recent papers \cite{ka1,ka2} I discussed the
  quantum statistics of the energy spectrum
   \bq E_n = \sigma \sqrt{n}E_P\;, n=1,2, \ldots,~~ E_P=
    c^2 \sqrt{c\, \hbar/G}~,~
   \sigma=O(1)~~, \eq whith
    degeneracies \bq d_n = g^n~,~g>1~~. \eq Many authors have proposed
   that  Schwarzschild black holes in 4-dimensional
     space\-times have such a spectrum (see the list in Ref.\ [1]).
     \\ The canonical
    partition function of the above spectrum not only leads
     to the Haw\-king temperature
    and the associated Bekenstein-Hawking entropy, but, very amazingly, it is
   formally the same as the {\it classical} grand canonical potential
     of the primitive droplet nucleation model in the context of first-order
     phase transitions in {\it two} space dimensions. Thus the so-called
     "holographic principle" \cite{tho1,tho2}, namely that the essential
 physics of
     black holes should be associated with the 2-dimensional horizon,
     is very evident here and comes out as a result! \\
     Furthermore, as the canonical partition function of the spectrum (1)
     becomes complex for the degeneracies (2), because $g>1$, one leaves
     the well-established frame\-work of equilibrium thermodynamics (KMS
     states)
     and
     moves on the perhaps more slippery ground of metastable states and
     nonequilibrium thermodynamics \cite{le1}.
 \\ The paper is organized as follows: I first
     generalize the  spectrum (1) to $ D \geq 4,$ spacetime
     dimensions \cite{ka3}. Then it will be shown that the resulting quantum
     canonical partition function is the same as the classical grand canonical
     potential of  the (nonrelativistic)
      primitive droplet nucleation model in $D-2$ spatial
     dimensions, the essential features of which for our purpose are discussed
     in chapter 3. \\ Using known results from the droplet nucleation model -
     especially the notion of a critical droplet - in chapter 4 the
     Hawking temperature and the Bekenstein-Hawking entropy are derived,
     up to a normalization factor which is  discussed separately. In chapter 5
     it is
     shown how an effective Hamiltonian of the Born-Infeld type can be
     used to describe  the spectrum plus
     the degeneracy factor. Its
     classical mechanical counterpart has some amusing properties, too.
     Finally, very preliminary and very sketchy, in chapter 6
      some surprising similarities to the very recent approach of
     understanding the quantum statistics of Schwarzschild black holes
     in the "Discretized Light-Cone Quantization"(DLCQ) version of
      "M(atrix)-theory" are
     pointed out which are perhaps not be accidental!
     \section{The quantum Schwarzschild BH spectrum  in $D$-dimensional
     spacetime} The spectrum (1) may be derived as follows: A canonical
     Dirac-type treatment of spherically symmetric pure Einstein gravity
     leads to a reduced 2-dimensional phase space having only the ADM mass
     $M$ and a canonically conjugate time functional $T$ as (observable)
     pair of variables \cite{ka4}. \\ An observer at spatially (flat) infinity
     will only have the mass $M$ and his own proper time $\tau$ available in
     order to describe the system. His very simple Schr\"odinger equation for
     it is
      \bq i\hbar \partial_{\tau} \phi(\tau) =Mc^2 \phi(\tau)~, \eq
which has the plane wave solutions \bq \phi(M,\tau)= \chi(M)e^{\ds
 -\frac{i}{\hbar}Mc^2\tau}~, \eq where $M \geq 0 $ is assumed. If the system
 with  mass $M$ stays there forever, then $M$ is to be considered as a
 continuous quantitiy. However -- just like the momenta of plane waves in a
 spatial box with finite extension --, if the above system, represented by the
 plane wave (4),  has only a finite
 duration $\Delta$ then this property may be (crudely) implemented by imposing
 periodic boundary conditions on the plane wave (4), implying the
 relation \cite{ka5}
  \bq c^2M\Delta = 2\pi \hbar~ n~,~ n=1,2,
 \ldots~.
   \eq  The question now is how to choose $\Delta$! As the only intrinsic
    quantity
   available at spatial infinity to characterize a
    time interval is $M$ itself, or
   a function of it, namely the Schwarzschild radius $R_S(M)$, we
   assume \cite{ka5}
   \bq \Delta=\gamma R_S(M)/c~, \eq
   where $\gamma$ is a dimensionless number of order 1. Inserting (6) into
   (5) gives the mass quantization condition
    \bq \gamma cM_n\, R_S(M_n) = 2\pi \hbar~ n~,~ n=1,2,
 \ldots~.
   \eq  For $D=4$ we have $R_S=2MG/c^2$ and the spectrum (1) results.
   However, I assume (7) to be valid in any dimension $D\geq 4$, because
   the Schr\"odinger Eq.\ (3) has to hold in any such dimension! \\
    Before I discuss
   the consequences let me point out that (7) is also the appropriate
   generalization of a Bohr-Sommerfeld type quantization of the 2-dimensional
   horizon as suggested very early by Bekenstein \cite{be1},
     Mukhanov \cite{mu1}
   (see also  Bekenstein's recent review \cite{be2}) and - in the context of
   string theory - by Kogan \cite{ko}: \\
   If we interpret $cM$ as canonical momentum and $R_S$ as canonical
   coordinate, then (7) is,  qualitatively,  nothing else but the
   old-fashioned quantum counting of phase space cells. \\ A relation
    like (7) appears
   also in the very recent discussion of the Schwarzschild black hole in the
   DLCQ-version of "M theory"\cite{ba1} (see ch.\ 6 below). \\
   In the following it is convenient to use these notations: We
   put $D=1+d=2+ \tilde{d}$: $d$ gives the number of space dimensions and
   $\tilde{d}$ the spatial dimensions of the black hole horizon. \\
   In D dimensions the Schwarzschild radius is given by \cite{pe1} \bq
   R_S(M) = \left(\frac{16 \pi G_D M}{c^2\,
    \omega_{\td}\;\td}\right)^{ 1/(\td-1)}~, \eq where $G_D$ is the
   gravitational
    constant in $D$-dimensional spacetime and $\omega_{\td} = 2 \pi^{(\td
    +1)/2}/ \Gamma( (\td +1)/2)$ is the volume of $S^{\td}$. Inserting
    this $R_S$ into the relation (7) gives the mass spectrum
    \begin{eqnarray} M_n & =& \sigma_D\, n^{ 1-\eta}\, m_{P,D}~,~~
     \eta=1/\td~, \\
   \sigma_D &=&  \left(\frac{(2\pi)^{\td -2}\;  \omega_{\td}\;\td}{8\,
    \gamma^{\td
    -1}}\right)^{\ds \eta}~, \nonumber \end{eqnarray} where \bq  m_{P,D} =
    \left(\frac{\hbar^{D-3} c^{5-D}}{G_D}\right)^{1/(D-2)}~,~~~
    l_{P,D}=\left(\frac{\hbar G_D}{c^3}\right)^{1/(D-2)} \eq are the
    corresponding Planck mass  and Planck length in D spacetime
    dimensions, respectively. \\
    As to the degeneracies I again assume (2) to hold. I would like to stress
    that this assumption is very important for the thermodynamics
    involved! Bekenstein \cite{be1,be2} and Mukhanov \cite{mu1} have presented
    convincing intuitive physical arguments for it, but it does not follow
    directly within the framework of Refs.\ [7] where the
 Schr\"odinger
    Eq.\ (3) comes from. Perhaps loop quantum gravity \cite{as1} or
    "Matrix-theory" (see below) can be of help here. Mukhanov uses information
    theoretical reasoning to put $g=2$. Bekenstein recently \cite{be2} has
    summerized the corresponding arguments. They are closely related to
    those of the "stretched horizon" approach by Zurek and Thorne \cite{zu}.
    Others \cite{tho2,so1} argue similarly. \\
    All those arguments are strongly supported by the results above: Combining
    the Eqs.\ (7) and (8) yields the quantized $\td$-dimensional
     horizon \bq A_{\td}(n)= (R_S(M_n))^{\td}\, \omega_{\td} =\frac{32\,
     \pi^2}{\gamma \td} \, n \, l_{P,D}^{\td}~~,~~ n=1,2, \ldots~. \eq
     Thus, the horizon is "paved" additively with Planck-sized area elements
    $l_{P,D}^{\td}$! \\ Such a pavement (parquet) can be assembled in many
    different ways, the total number of possibilities being $2^{n-1}$.
    The ansatz (2) is a mathematically convenient generalization of this! \\
    Another point to comment on is the free parameter $\gamma$.
     We shall see below how
   the Gibbons-Hawking geometrical approach \cite{gi1} to the
     partition function may be used to restrict it, but it is
      worthwhile to point out
     that such a free parameter also occurs in loop quantum gravity \cite{as1}
     and in the discussion of the Schwarzschild black hole thermodynamics
     of  Matrix theory \cite{ba1}. \\ The canonical partition function can now
     be written as
     \begin{eqnarray} Z_D(t,x)& =&
   \sum_{n=0}^{\infty}e^{\ds n t -n^{1-\eta}x}~, \\
  & &t= \ln g~,~~ \eta = 1/\td \equiv 1/(D-2)~, \nonumber \\
  & &  x= \beta \,\sigma_D\, E_{P,D}~,~~\beta = \frac{1}{k_B T},~~
  E_{P,D}=m_{P,D}\,
  c^2 ~. \nonumber \end{eqnarray}  $Z_D(t,x)$ obeys the linear PDE
 \bq \partial_t^{\td-1}Z_D = (-1)^{\td} \,\partial_x^{\td}
    Z_D, \eq which can be used to determine $Z_D$ in closed form. This has
    been done \cite{ka6} for $\td=2$ and $3$.
  \section{Primitive droplet nucleation model   in $D-2$ space
 dimensions}
  The model assumption that 1st-order phase transitions
   are  initiated by the
 formation (homogeneous nucleation) of expanding droplets
  of the new phase within the old phase is a popular and important
  one (see the reviews \cite{rev} with their references to the original
  literature). In its most primitive form the droplets are
  assumed to be spherical and  to
  consist of $n$ constituents (e.g.\ droplets of Ising spins on a lattice or
  liquid droplets of molecules etc.),
  the "excess" energy $\epsilon_n$ of which is given by a  "bulk"
   term proportional to the volume
  $n$ and a  term proportional to the surface $n^{1-\eta}, \eta =1/\td$, where
  $\td \ge 2$ is the spatial dimension of the system:  \bq   \epsilon_n =
   -\hat{\mu}\, n +
  \phi \, n^{1-\eta}~,~~\eta =1/\td~.\eq In the case of negative Ising spin
  droplets, formed in a background of positive spins by turning
  an external magnetic field
  $H$ slowly negative - below the critical temperature -, the coefficient
  $\hat{\mu}$ in Eq.\ (14) takes the form $\hat{\mu}=-2H$ and for liquid
 droplets of $n$
  molecules condensing from a supersaturated vapour one has $\hat{\mu}= \mu
  -\mu_c$, where $\mu$ is the chemical potential and $\mu_c$ its critical
  value at condensation point. $\phi$ is the constant surface energy.\\
   Assuming the average number $\bar{\nu}(n) $ of
  droplets with
  $n$ constituents to be proportional to a Boltzmann factor, \bq
    \bar{\nu}(n)
 \propto e^{\ds -\beta
  \epsilon_n},~ \beta = \frac{1}{k_B T}~, \eq and  that the droplets form
  a noninteracting dilute gas leads to the grand canonical
   potential $\psi_{\td} $
  per spin or per
  volume \begin{eqnarray}
  \psi_{\td} (\beta, t = \beta \hat{\mu}) & =& \ln Z_G  = p \beta =
   \sum_{n=0}^{\infty}e^{\ds  t n-x n^{1-\eta}}~, \\ & & t=\beta
   \hat{\mu},~ x=\beta \phi~;~ p:\mbox{pressure}, \nonumber \\ & &
 d\psi_{\td} = -U d \beta +
    \bar{n}dt~~ . \end{eqnarray}
  (For physical reasons  the sum (16) may not start at $n=0$ but at some
  finite $n_0 >0$. This can easily be taken care of.
  It is mathematically convenient to start at $n=0$.) \\
  Obviously the sum (16) is the same as in (12)! \\ The interpretation of the
  sum (16) as a grand canonical potential comes about as follows
  \cite{la1,po1}:
  One starts with a canonical partition function $Z_c$, where the same terms
  as above are summed up. However, then one has the thermodynamics of
  single $n$-droplets, but there may be $N$ of them. If one assumes these to
  be noninteracting and indistinguishable, then the grand canonical partition
  function is \bq Z_G= \sum_N \frac{Z_c^N}{N!}= e^{\ds Z_c}~, \eq which leads
  to the grand canonical potential $\psi_{\td}$ of Eq.\ (16). \\ It will be
  important, however, that we interpret $Z_D$ of (12) as a {\it canonical}
  partition
  function, where $g=e^t$ describes the  fixed $temperature-independent$
  degeneracies
  of the corresponding quantum  levels, whereas
   in the droplet nucleation model $z=e^t, t= \hat{\mu}\beta$,
  is  the $temperature-dependent$ fugacity
 of a classical
  Boltzmann gas! \\  Notice that $\psi_{\td}$ contains no explicit information
  about properties of the phases before and after the phase transition.
  Consider, e.g., a vapour $\rightarrow$ fluid phase transition. Then the
  properties of the vapour  are only very indirectly present in $\psi_{\td}$,
  namely in form of the
  surface energy $\phi$ of the droplets emerged in the vapour. The model
 here merely is supposed (for more details see the Refs.\ [20])
  to describe that  (metastable!)
 part of the
  Van der Waals isotherm
 in the $(V,p)$-plane which starts where, with decreasing volume,
 the  (theoretically!) strict equilibrium
 line of the Maxwell construction branches off to the left,
  till the (local) maximum of the Van der Waals "loop", the "spinodal" point,
 is reached.
  \\ The series (16) converges for $t\le 0$ only.  This follows, e.g., from
the  Maclaurin-Cauchy integral criterium \cite{14}.
 In applications to metastable systems, however,
  one is
  interested in the behaviour of $\psi_{\td}(t,x)$ for
   $t \geq 0$. This calls for
  an analytic continuation in $t$ or in the fugacity $z=e^t$ which reveals
  a branch cut of $\psi_{\td}$ from $z=1$ to $z= \infty$ \cite{br}. \\
Qualitatively the following happens: For $t<0$ (i.e.\ positive magnetic field)
$\epsilon_n$ increases monotonically with $n$, making the corresponding terms
in $\psi_{\td}$ decrease monotonically. The small droplets are favoured and
no phase transition occurs. \\ If, however, $t>0$, then $\epsilon_n$ has a
maximum for \bq n^*=\left(\frac{(1-\eta)\phi)}{\hat{\mu}}\right)^{ \td} =
         \left(\frac{(1-\eta)x}{t}\right)^{ \td}~,~x=\phi\beta~, \eq with
      \bq \epsilon^* \equiv \epsilon_{n^*} =a \eta (1-\eta)^{\td -1}~,~~
      a=\frac{\phi^{\td}}{\hat{\mu}^{\td-1}}=\frac{x^{\td}}{\beta
 t^{\td -1}}~, \eq
      after which $\epsilon_n$ becomes increasingly negative with increasing
      $n$ and the series (16) explodes! \\ The physical interpretation
       is the
      following: If, by an appropriate fluctuation, a "critial droplet" of
      "size" $n>n^*$ has appeared, it is energetically favoured to grow. Such
 an  overcritical droplet
      -- and others of a similarly large size -- will destabilize the
 original phase and
      will send the system to the phase for which it has served as a
 nucleus! \\
      The energy $\epsilon^*$ may be interpreted as a measure for
      the critical barrier of the
      free energy the system has to "climb" over in order to leave the
 metastable
      state for a more stable one. \\
      Furthermore, the rate $\Gamma$ for the transition of the metastable
       state to the more stable one is
      proportional to $\exp(-\beta \epsilon^*)$. However, calculating the
      rate is no longer a problem of {\it equilibrium} thermodynamics. One
 has to
      deal with tools of {\it nonequilibrium} processes like the Fokker-Planck
      equation etc. For the droplet model this was essentially
       done by Becker and D\"oring \cite{be}. They
       assumed a stationary situation, where a steady flow of small,
      but in size increasing, droplets leave the metastable state and all
     overcritical droplets which have passed the barrier are removed from the
      system. \\Their approach was considerably improved by Langer \cite{la2}
       who related
      the transition rate $\Gamma$ to the imaginary part of $\psi_{\td}$.
      This can be seen roughly as follows:
 If one turns the sum
  $\psi_{\td}$ in (16) into an integral by interpreting  $n$  as a continuous
   variable:
   \bq \tilde{\psi}_{\td}=
  \int_0^{\infty}dn e^{\ds \beta(\hat{\mu} n -\phi\, n^{1-\eta})}~~, \eq
 then a saddle point approximation [21,19] for large $\beta$ gives the
 asymptotic expansion
 \bq \tilde{\psi}_{\td} \sim
  (1-\eta)^{\td/2}\sqrt{\frac{\pi\, \td}{2 \hat{\mu}\, \beta}} \left
  (\frac{\phi}{\hat{\mu}} \right)^{\td/2}
  e^{\ds -\beta a \eta(1-\eta)^{\td-1}}(i+O(1/\beta))~. \eq Here the path
 in the complex
  $n$-plane goes from $n =0$ to $ n^*$ and then parallel to the imaginary
  axis to $+i\infty$ \cite{ka6}. (Thus, only half of the associated Gaussian
 integral
  along the steepest descents contributes!) \\ The crucial point is that
  the saddle point is given by $n^*$ and the associated $\epsilon^*$ of
  Eqs.\ (19) and (20), that is to
  say, by the critical droplet! \\ The leading term in the saddle point
  approximation (22) is purely imaginary. Performing a Fokker-Planck type
  analysis, Langer found \cite{la2} that the transition rate $\Gamma$
  is essentially proportional to the imaginary part
 $\Im (\tilde{\psi}_{\td}),$ the other factor
  being a "dynamical" one, genuinely related to nonequilibrium properties. \\
  An essential point for us here is the result that the imaginary part
  of $\psi_{\td}$ can be interpreted, at lest intuitively,
   in terms of equilibrium concepts although
  it is related to nonequilibrium properties which are, however, near to
  stationary situations. For further discussions see the reviews mentioned in
  Ref.\ [20].
 \section{Hawking temperature  and Bekenstein-Hawking entropy}
 We are now ready to apply the droplet nucleation model results to
 the Schwarzschild black hole: If we denote the "critical" term in the series
 (12) by $Z^*_{D}$,  we have   \bq Z^*_{D}= e^{\ds -
 [\eta(1-\eta)^{\td-1}x^{\td}]/t^{\td-1}}~. \eq The essential point now is that
 $t=\lng$ here is no longer a temperature dependent quantity -- as in the
 droplet model -- but a fixed number. Therefore the equation for the associated
 internal
 energy, \bq U^*=-\frac{\partial \ln Z^*_D}{\partial \beta}=
 (1-\eta)^{\td-1}\left(\frac{x}{t}\right)^{\td-1} \sigma_D E_{P,D}~ = \td\,
 \epsilon^* ~, \eq
 can be used to determine  the (inverse) temperature $\beta^*$ needed
for a (potential) heat bath, if  the rest energy
 $U^*$ is given! Solving Eq.\ (24) for $x$ and using the relation (8) between
 Schwarzschild radius $R_S$ and mass $M=U^*/c^2$ we obtain \bq
\beta^*=
 \lambda \left(\frac{4\pi
R_S^*}{(\td-1)\hbar c}\right)~,~~ \lambda \equiv \frac{t\, \td\,
 \gamma}{8\pi^2}~,
\eq where $R_S^*=R_S(M=U^*/c^2)$ (Eq.\ (8)). \\ The expression
in the bracket of Eq.\ (25) is exactly the inverse Hawking temperature
in $D$-dimensional spacetime \cite{pe1}, if we
 identify $U^* = Mc^2$, where $M$ is the
macroscopic rest mass of the black hole! Thus, up to a numerical
factor $\lambda$ of order 1, we obtain the Hawking
 temperature in this way. \\ For the
entropy $S^*_D=\beta^* U^* + \ln Z^*_D$ we get \begin{eqnarray}
 S^*_D/k_B
 &=& (1-\eta)^{\td}
(x^*)^{\td}/t^{\td-1}= (1-\eta)\beta^* U^* \\ & =&
 t\, n^* = \ln (g^{\ds n^*})~,~~ x^*=\beta^*\sigma_D E_{P,D}~, \nonumber
  \end{eqnarray} where
$n^*$ is the same as in (19). \\
If we express $S^*_D$  in terms of the $\td$-dimensional surface $A_{\td} =
\omega_{\td}\;(R_S)^{\td}$, we have \bq S^*_D/k_B = \lambda\;
\frac{A_{\td}}{4 l_{P,D}^{\td}}~ = \lambda\;
\frac{c^3\,A_{\td}}{4 \hbar G_D}~~ . \eq
 So, up to the same numerical factor
$\lambda$ we  already
encountered in connection with the inverse temperature,
 we obtain the Bekenstein-Hawking
 entropy! \\
  The mean square fluctuations of the energy \bq
 (\Delta E)^2 = \partial^2_{\beta}(\ln Z^*_D)=-
 \frac{(1-\eta)^{\td-1}(\td-1)}{t^{\td-1}}\;
  x^{\td-2}\; (\sigma_D E_{P,D})^2 \eq
 are negative (negative specific heat!), but relatively small for large masses
 because \bq \frac{(\Delta E)^2}{U^*}=-\frac{\td-1}{\beta^*}~~, \eq
 which appears to be quite an universal relation: the r.h.s.\ of Eq.\
 (29) depends only on $\td$ and $\beta^*$! \\
 As the factor $\gamma$, up to now,  is a free parameter, we can possibly
choose it in such a way that
the above prefactor $\lambda$ equals one: \bq \gamma= \frac{8 \pi^2}{
t\,\td}~.
 \eq A suitable argument for such a normalization $\lambda=1$
comes from the practically classical geometrical result for $S$ of
 Gibbons and Hawking \cite{gi1}, derived from the euclidean section of the
 Schwarzschild
 solution. In view of the surprising approximate equalities of the classical
 and the quantum theoretical values for $S$, one may use the classical result
 for normalizing the quantum one. \\ There is  a corresponding analogue in QED,
  where the physical value
 of the electric charge $e$ in the quantum theory (to all orders) is
 normalized via
  the universal classical total
 Thomson  cross section $\sigma_{tot} =(8/3)\,\pi r_0^2, r_0= e^2/(mc^2),$
  for Compton scattering in the limit of vanishing photon energy
   \cite{qed}. \\
  One has to be
 careful here, however, because only $Z^*_D$ has the same simple exponential
 form as one finds in the Gibbons-Hawking approach in lowest order.
 Fluctuations lead to  prefactors with powers of $x$ in front of $Z^*_D$
 as can already be seen if we take
 the imaginary part $ Z_{i,D}$ of the saddle point approximation (22) as
a slightly more sophisticated "pseudo" partition function, or, if we take
the imaginary
part of the purely imaginary partition function for the (euclidean)
Schwarzschild black hole if one includes ``quadratic'' quantum fluctuations
around the classical solution \cite{gi2,pe2,ka2}.
 Such additional powers of $x$
lead to corrections to $\beta$ and to
 logarithmic corrections of the entropy (27)
\cite{ka1}. \\ As to the  connection between the statistically defined
entropy and the one obtained geometrically   and as to possible quantum
 corrections
see the recent review by Frolov und
Fursaev \cite{fr1}. \\ The corresponding normalization problem in loop
quantum gravity has been discussed in Refs.\ \cite{as1,ro}. \\ I
 said "pseudo"
 partition function because they imply negative
mean square fluc\-tu\-ati\-ons, see Eq.\ (28) and Ref.\ \cite{ka1}, which
 appears to be associated
 with the
metastability of the system. More theoretical work seems to be
 necessary in order to
understand these partially surprising features better beyond the realm of
strict equilibrium thermodynamics. \\ Altogether we see the following picture
emerging: If the quantum system in $D$ spacetime dimensions of (whatever)
total mass $M$
and  represented by the plane wave (4)  collapses
after a finite time, then the theoretical implementation of
 this collaps through
periodic boundary conditions enforces the mass quantisation  (7), which may
also be obtained  heuristically through the old-fashioned Bohr-Sommerfeld
 rules. \\ If, in addition, the degeneracies (2) are assumed, the
 associated quantum statistics is formally  the  same as that of the
classical primitive droplet nucleation model in $D-2$ space dimensions,
 represented as a classical grand canonical ensemble.
 This shows how the
 holographic principle is indeed implemented. It  further indicates how the
 quantum background is hidden behind a classically appearing facade, very
 probably formed
 by the thermal physics of the horizon, which is being built up during the
  nucleation of the black hole. \\ The "blurring" of the quantum properties
  is also indicated by the fact that the
  traces of the
  Bose statistics of the quanta (1) are rather hidden,
  contrary to, e.g., the canonical
  partition function of the simple harmonic oscillator. \\ If we look at the
   exact expression for $Z_4$ in closed form, derived in Ref.\ [1],
  \begin{eqnarray} Z_4(t, x)&=&
\int_0^{\infty}d\tau
  \hat{K}(\tau, x)\frac{1}{1-e^{\ds (t-\tau)}}~,~\\ & &
\hat{K}(\tau,x) =
   \frac{x}{2\sqrt{\pi \tau^3}}e^{\ds -x^2/(4\tau)}~, \nonumber \\
      \Re [Z_4(t,x)]&= &\mbox{p.v.} \int_0^{\infty}d\tau\,
      \hat{K}(\tau,x)
   \frac{1}{1-e^{t-\tau}}~,~~
\Im [Z_4(t,x)] = \pi
  \hat{K}(t,x)~,
  \nonumber \end{eqnarray} only the factor $1/(e^{t-\tau}-1)$
   in the principal
 value integral for the {\it real} part of $Z_4$ indicates Bose statistics,
 whereas the, for our discussion above crucial, imaginary part does not show
  such
traces (see also the next chapter). Here may lie a key to the information
loss problem \cite{ha}.\\
  The role of the external magnetic field (or a corresponding chemical
  potential) in the droplet case  finds its correspondence in the degeneracy
  factor $g$ which intuitively represents the gravitational pull leading
  to the formation of the black hole by nucleation. \\ What is really
  new, compared to the droplet model,
   is that the free energy barrier $\epsilon^*$ determines its own
  temperature, namely $T_H$, and the associated entropy (26),
   as a function of the total internal energy $U^* =Mc^2$.
   This is a genuinely quantum mechanical effect,
  because $t= \ln g$ is a fixed number, not a temperature-dependent quantity
  as in the droplet nucleation case. Here lies the real difference.
  This temperature, after nucleation, then becomes that of the black hole
  itself. Any thermal radiation emitted from the horizon carries the imprint
  of this temperature. \\ In the above physical interpretation of the
 nucleation process concerning the black hole I followed the droplet nucleation
 picture and have assumed that the  black hole is the $result$ of the decay
 of metastable states to a more stable one (the black hole) which
 then Hawking radiates with the corresponding temperature. \\ Another
interpretation is that the black hole itself is the metastable state which
 slowly decays, due to its interaction with the heat bath consisting of
 Hawking radiation. A decision between these two alternative pictures probably
 needs the explicit introduction of matter fields coupled to the quantum
 black hole
in order to see what really happens. \\
 The nucleation of black holes has been discussed
  quite early by Gross, Perry and Yaffe in terms of the euclidean Schwarzschild
  instanton \cite{pe2,ka2}. \\ One can arrive at similar results for the
 Hawking
  temperature and the Bekenstein-Hawking entropy as above if one performs a
  microcanonical counting of states \cite{sch}, however, then one
   loses the very
  inspiring connection to the droplet nucleation picture.
  \section{Effective quantum  and  classical Hamiltonians} The partition
  function $Z_D$ of Eq.\ (12) may be rewritten as \bq Z_D = \mbox{tr}(e^{\ds -
  \beta \hat{H}})~,~\hat{H}= -\mu a^+a + \epsilon (a^+a)^{1-\eta},~
   \mu=t/\beta,~
  \epsilon = \sigma_D E_{P,D}~, \eq where $a$ and $a^+$ are the annihilation
  and creation operators of the harmonic oscillator. If $u_n$ is an
  eigenfunction of the harmonic oscillator, then $a^+a\,u_n = n\, u_n$ and the
  assertion (32) follows immediately. As the trace is independent of the basis
  one uses for its calculation, one might also use another one, e.g.\ the
  coherent states $|z>$, which are eigenstates of $a$ with complex eigenvalues
  $z$. Doing so \cite{lo} for $\td =2$ leads to the same exact result as in
  Ref.\ [1]. \\ It is amusing to look \cite{mu3} briefly at the corresponding
 classical
       effective system: Let us define the classical
       quantity \bq \tilde{N}=\frac{1}{\hbar \omega_0}
       \left(\frac{1}{2m}p^2+\frac{m}{2}\omega_0^2 q^2\right)~. \eq  After an
       appropriate rescaling of $q$ and $p$ we have the effective Hamiltonian
  \bq \tilde{H} = -\mu \tilde{N}+ \epsilon
  \tilde{N}^{1-\eta}~,~~\tilde{N}=\frac{1}{2}(p^2+q^2)~,\eq leading to
  \begin{eqnarray}\dot{p}& =& - \frac{\partial \tilde{H}}{\partial q}=
  -\tilde{\omega}q~,~~\tilde{\omega}(\tilde{N})=-\mu+(1-\eta)\epsilon
 \tilde{N}^{\ds
  -\eta}~~, \\ \dot{q}&=& \frac{\partial \tilde{H}}{\partial p}=
  \tilde{\omega}p~. \end{eqnarray} It follows that $\tilde{N}$ is a constant
  of motion $ \tilde{N}_0$ for the associated fictitious point particle
   and that this
  particle moves on a circle with radius $\sqrt{2\tilde{N}_0}$ in phase space
 with frequency $\tilde{\omega}$
  which is a function of $\tilde{N}_0$ (this is a new feature
  compared to the usual harmonic oscillator). \\ The critical value
  $\tilde{\omega}= 0$ results if $\tilde{N}=\tilde{N}_c=((1-\eta)
\epsilon/\mu)^{\td}$
  which is just the same as $n^*$ from Eq.\ (19) above and for which
 $\tilde{H}$ has its
  maximum. For $\tilde{N}<\tilde{N}_c$ the frequency $\tilde{\omega}$ is
 positive and
  for $\tilde{N}>\tilde{N}_c$ it is negative. \\ As \bq \dot{q} =
  ((1-\eta)\epsilon \tilde{N}^{-\eta} -\mu)p~, \eq it is in general not at
  all trivial to calculate the Lagrange function $L(q,\dot{q})$, because one
  has to solve an algebraic equation if one wants $p(\dot{q})$ from  (37).
  Already for $\td =2$ this equation is of order 4.
   If $\mu =0$ we get in this case
  \bq L(q,\dot{q})=-q(\frac{1}{2}\epsilon^2-\dot{q}^2)^{1/2}~, \eq
  which may be interpreted as a simple example of a Born-Infeld type
  La\-grange\-an \cite{bor}. \\ Finally I mention how the classical partition
 function
 $Z_{cl}$ associated with the Hamiltonian (34) looks like: \bq
 Z_{cl}=\int_{-\infty}^{\infty}\int_{-\infty}^{\infty}
\frac{dpdq}{2\pi \hbar} \exp\{-\beta
[-\frac{\mu}{2
 \hbar}(p^2+q^2)+\frac{\epsilon}{\sqrt{2\hbar}}(p^2+q^2)^{1/2}]\} ~. \eq
 Introducing polar coordinates in the $(q,p)$-plane and making appropriate
 substitutions gives \bq Z_{cl}= 2\int_0^{\infty}du u e^{\ds t u^2-xu}~,~
 t=\beta \mu,~x=\epsilon \beta~~. \eq The integral exists for $t<0$ and
 gives \cite{GR1} \bq Z_{cl}=-\frac{e^{\ds
 -x^2/(8t)}}{t} D_{-2}(\frac{x}{\sqrt{-2t}})~,\eq where $D_p(z)$ is the
 parabolic
 cylinder function of order $p$. Continuing now from  negative to
 positive $t$ finally yields \bq
 Z_{cl}=-\frac{1}{t}\Phi(1,1/2;-\frac{x^2}{4t})+i\frac{\sqrt{\pi}x}{2t^{3/2}}
 e^{\ds -x^2/(4t)}~, \eq where $\Phi(a,c;z)$ is the confluent hypergeometric
 function \cite{er} with $\Phi(a,c;z=0)=1$. \\ What is remarkable
 here is that the imaginary part
 of the classical partition function $Z_{cl}$ is the same as that of the
 quantum theoretical one (31). The real parts are different (put $x=0$),
 however, reflecting the difference between classical and quantum mechanics.
 \\ The "invariance" of the imaginary part under quantization reminds one of
  the Rutherford scattering cross section for charged particles which is
  the same in mechanics and quantum mechanics,
   due to the long range of the Coulomb forces.
\section{Remarks on possible relationships to DLCQ Matrix Theory}
 The following
remarks are very preliminary and very qualitative , but I hope
they are nevertheless useful and
not too much beside the point! They are made under the assumption
that the correspondence between the quantum statistics of Schwarzschild black
holes and the classical droplet nucleation model is not accidental. \\
Let me start by slightly changing the language: Instead of a ``droplet'' in
$\td$ space dimension we may speak of a spherical ``$\td-brane$'' with an
associated ``$(\td-1)$-brane'' as its boundary. For $D=4$ the corresponding
objects in 2 dimensions are compact membranes (``2-branes'') with  closed
strings as their boundary. Recall that closed strings are essential for
having gravity in an effective low energy string limit \cite{wi}. If the
$\td$-branes
 are
spherical they are in their ground states. \\ For $n<n^*$ the
 $\td$-branes are
(meta)stable, but for $n>n^*$ they are unstable, nucleating the new
phase, the black hole. \\ This role of the critical droplet size $n^*$
finds its analogy in the "correspondence principle
for black hole and strings" \cite{su1,ho1} which says that a highly excited
string state becomes a black hole when its length scale $l_S$ shrinks to
 less than the Schwarzschild radius $R_S$ of the associated black hole of the
 same mass. At this "correspondence point", which can be reached
by adjusting the string coupling constant $g_S$, the entropies
 of the two systems
become comparable if their masses and charges are. This then opens the
 possibility to
calculate the entropy of black holes by counting the states of an
associated string system. \\ For Schwarzschild black holes this program
has very recently been discussed in the framework of the "DLCQ"-version
of "Matrix theory" (as to this see the reviews \cite{ba2}) which
structurally - not in detail, of course - shows some surprising
 similarities with the droplet nucleation picture which are, perhaps,
not  accidental: \\ In the DLCQ approach (in the following natural units
 are beeing used) one compactifies the lightlike
 coordinate
$x^- = x^0-x^{10}$ of the 11-dimensional theory on a circle of
 radius $R$, thus
making the related momentum discrete, $P_- =N/R, N=1,2,\ldots$,
 where $N$ labels
the different "parton" sectors in the infinite momentum frame. The number of
states belonging to a given $N$ depends on the supersymmetric Galilei
 invariant part of the system in the
 nine "transverse" directions  several of which may be compactified. \\
 The number $N$ associated with a $lightlike$ compactification corresponds
 to the number $n$ of Eqs.\ (5) or (7) coming from a $timelike$
 compactification.
 This correspondence  be\-comes evident as follows: in the DLCQ framework
 one  asks what is
 the minimal $N=N_{min}$ such that an appropriate Lorentz boost makes $R$
  large enough for a black hole of mass $M$ just to  fit in. The answer
is \bq N_{min} \approx M R_S \approx S~,\eq where $S$ is the entropy.
Thus, it is  $N_{min}$
which
characterizes the correspondence point \cite{ba1,ho2,da}! Obviously
$N_{min}$ plays quite a similar role as the critical $n^*$ in the
 droplet model above
 (Eqs.\
(7) and (26)). \\The
counting of states which yield the Bekenstein-Hawking entropy and the Hawking
temperature are, of course, very different: in our case the essential input is
(2), whereas in the DLCQ approach one either counts the states from the SUSY
YM fields etc.\ just {\it below} the correspondence point \cite{ba1,ho2,en}
 or the states of the
nonrelativistic D0-brane gas slightly {\it above} that
point \cite{ho2,ts}. In the latter case one needs Boltzmann counting
  which one
obviously has in the droplet nucleation model, too. \\A very essential
difference  between the DLCQ approach and the droplet nucleation model is,
of course, that the latter is  not
supersymmetric. \\  Questions are: Can Matrix theory (or string theory)
determine the normalization factor $\lambda$ of Eq.\ (27) for neutral
 Schwarzschild
black holes (see Refs.\ \cite{en}), can it provide the spectrum (9) (very likely
in view of Eq.\ (43)) and, most importantly, can it explain the degeneracies
(2) and the value of $g$?
\section{Acknowledgments} I thank Jacob Bekenstein for encouragements,
Slava Mukhanov for an invitation to the ETH Z\"urich and for stimulating
discussions, Don Marolf for inviting me to give a talk at the
GR 15 conference in Pune. I had a number of stimulating discussions at
that conference, in particular with Valery Frolov. \\ Especially I thank
 Sumit Das for his invitation
to stay afterwards for a week  with the Theoretical Physics Group of the
Tata Institute in Bombay. I enjoyed its  friendly and warm
hospitality very much
and I thank Sumit Das and P.\ Ramadevi for encouraging and helping my
 beginning attempts
to understand the essential features of Matrix theory. I thank T. Strobl for
 a critical reading of the manuscript. Last, but not least,
 unlimited thanks to my wife Dorothea!


\begin{thebibliography}{99}
\bibitem{ka3} This possibility was already indicated in a
remark added at the end of Ref.\ [2], as an answer to a question of an
anonymous referee, whom I thank for that question.
\end{thebibliography}
 \end{document}